\documentclass[aps,prb,superscriptaddress,showpacs,byrevtex,twocolumn]{revtex4-1}

\usepackage{color}

\usepackage{graphicx}
\usepackage{amsmath}
\usepackage[english]{babel}
\usepackage{amsfonts}
\usepackage{amssymb}
\usepackage{siunitx}
\usepackage[latin1]{inputenc}
\begin{document}

\title{Interface originated modification of electron-vibration coupling in resonant photoelectron spectroscopy}
\date{\today}
\author{C. Sauer}
\author{M. Wie{\ss}ner}
\author{A. Sch\"oll}
\author{F. Reinert}
\affiliation{Universit\"at
W\"urzburg, Experimentelle Physik VII \& R\"ontgen Research Center
for Complex Material Systems RCCM, 97074 W\"urzburg, Germany}
\affiliation{Karlsruher Institut f\"ur Technologie KIT,
Gemeinschaftslabor f\"ur Nanoanalytik, 76021 Karlsruhe, Germany}


\begin{abstract}
We present a comprehensive study of the photon energy ($h \nu$) dependent line-shape evolution of molecular orbital signals of large \mbox{$\pi$-conjugated} molecules by resonant photoelectron spectroscopy (\mbox{RPES}). A comparison to \mbox{RPES} data of small molecules suggests that the excitation into different vibrational levels on the intermediate state potential energy surface of the electronic excitation is responsible for the observed effect. In this simplified picture of electron-vibration couping the character of the potential energy surfaces involved in the \mbox{RPES} process determines the line-shape of the molecular orbital signal for a particular $h \nu$. We use the sensitivity of this effect to probe the influence of different interfaces on the electron-vibration coupling in the investigated systems. The magnitude of the variation in line-shape throughout the particular $h \nu$ region allows to reveal significant differences within the physisorptive regime.
\end{abstract}

\maketitle


\section{Introduction}
A generally interesting aspect in the description of quantum mechanical systems is the coupling of different excitations to one another. A prominent example is electron-vibration coupling i.e. the excitation of vibrations accompanied to electronic excitations. Within the Born-Oppenheimer approximation and by the Franck-Condon principle a fairly simple and intuitive picture of electron-vibration coupling in photoelectron spectroscopy (PES) \cite{Huefner} and near edge X-ray absorption fine structure (\mbox{NEXAFS}) spectroscopy \cite{Stoehr} is achieved. This allows to obtain information about electron-vibration coupling by analyzing the line-shape of PES spectra of large \mbox{$\pi$-conjugated} molecules and thereby extracting a crucial parameter for charge transfer \cite{UenoProgSurfSci,KeraProgSurfSci} in application relevant systems \cite{HwangMatSciEn}. With a line-shape analysis of \mbox{NEXAFS} spectra of such molecules parameters of the core excited potential energy surface can be obtained \cite{SchoellPRL}. In special cases an electron-vibration coupling investigation of \mbox{NEXAFS} line-shapes can even serve as a probe for electronic coupling of molecules \cite{ScholzPRL}. Due to the surface sensitivity of PES and partial electron yield \mbox{NEXAFS} these techniques are well suited to study thin films and interfaces. Furthermore \mbox{NEXAFS} excitations at different elements or different chemical species of the same element can provide site specific information for homo-molecular systems. For hetero-molecular systems \cite{BobischJChemPhys,ChenApplPhysLett,KasemannLangmuir,HaemingPRB,HaemingSurfSci,StadtmullerPRL} the excitation can be chosen to obtain molecule specific information.

Resonant PES (\mbox{RPES}) allows to take the investigation of electron-vibration coupling to a next level. \mbox{RPES} experiments on small molecules (CO, N$_2$, O$_2$) in the gas phase in comparison to calculations for example reveal that lifetime vibrational interference \cite{NeebJourElSpec,OsborneJChemPhys} or $h \nu$ detuning \cite{SundinPRL} can significantly alter the line-shape of molecular orbital derived signals which is determined by the electron-vibration coupling. Using state of the art experimental equipment in combination with theoretical analysis even imaging of the ionic molecular potentials and mapping of vibrational wave functions is feasible for a simple molecule as N$_2$ \cite{KimbergPRX}. 

A comparable \mbox{RPES} investigation of the intensively studied large \mbox{$\pi$-conjugated} molecules leads to a dramatic increase of complexity. Many close lying electronic excitations in the \mbox{NEXAFS} spectrum and the possibility of multiple vibronic modes coupling to each of them and to each other poses a tremendous challenge for an investigation of electron-vibration coupling. However, the application of this powerful technique to such complicated systems does not only raise the level of complexity but in fact allows to obtain additional information with respect to PES or \mbox{NEXAFS} alone, while simultaneously retaining the advantages of both. A selective excitation into highly excited vibrational levels $\nu_i$ in the intermediate state leads to a line-shape of the \mbox{RPES} signal from molecular orbitals which is very sensitive to the character of the involved potential energy surfaces $V$ (ground state $V_g$, intermediate state $V_i$ and the final state $V_f$). Thus in principle \mbox{RPES} is able to investigate electron-vibration coupling in the final as well as in the core excited intermediate state if the line-shape of a proper signal can be analyzed in sufficient detail. 

In this work we want to push the investigation of electron-vibration coupling in large \mbox{$\pi$-conjugated} molecules forward by performing a large comprehensive study on the $h \nu$ dependent evolution of the line-shape of molecular orbital signals in \mbox{RPES}. This previously observed effect \cite{KjeldgaardPRB,HaemingChemPhysLett} is first discussed for the signal originating from the highest occupied molecular orbital (HOMO) of a copper-phthalocyanine (CuPc) multilayer and then related to \mbox{RPES} investigations of small molecules. This comparison is used to develop a simplified picture in which the character of the involved potential energy surfaces are the only necessary parameters to consistently discuss the observed line-shape evolutions. Both the discontinuous line-shape evolution of coronene and the continuous line-shape evolution of CuPc and tin-phthalocyanine (SnPc) in the energy region of interest are related to examples of small molecules. The parallel but \mbox{NEXAFS} resonance specific behavior of the signals from the three frontier molecular orbitals of coronene constitutes an additional successful consistency check of the applied picture. Moreover the sensitivity of the technique is revealed by a comparison of CuPc and SnPc multilayer films and then used to probe the influence of different interfaces. It is found that the adsorption on Au(111) influences the line-shape evolution of the HOMO signal of CuPc and SnPc differently in contrast to the similar modification observed at a hetero-organic interface. In the former case less relative intensity in high final state vibrational levels $\nu_f$ with respect to the multilayer is observed for CuPc on Au(111) while SnPc exhibits no significant alteration. In the later case both systems show an interface induced enhancement of relative intensity distribution into higher $\nu_f$. Finally we present a resonance specific interface modification of the $h \nu$ dependent line-shape evolution by the adsorption of coronene on Ag(111). All observed changes can be consistently discussed on the basis of an influence on the character of the potential energy surfaces involved in the \mbox{RPES} process.

\begin{figure*}
    \centering
        \includegraphics[width=0.9\textwidth]{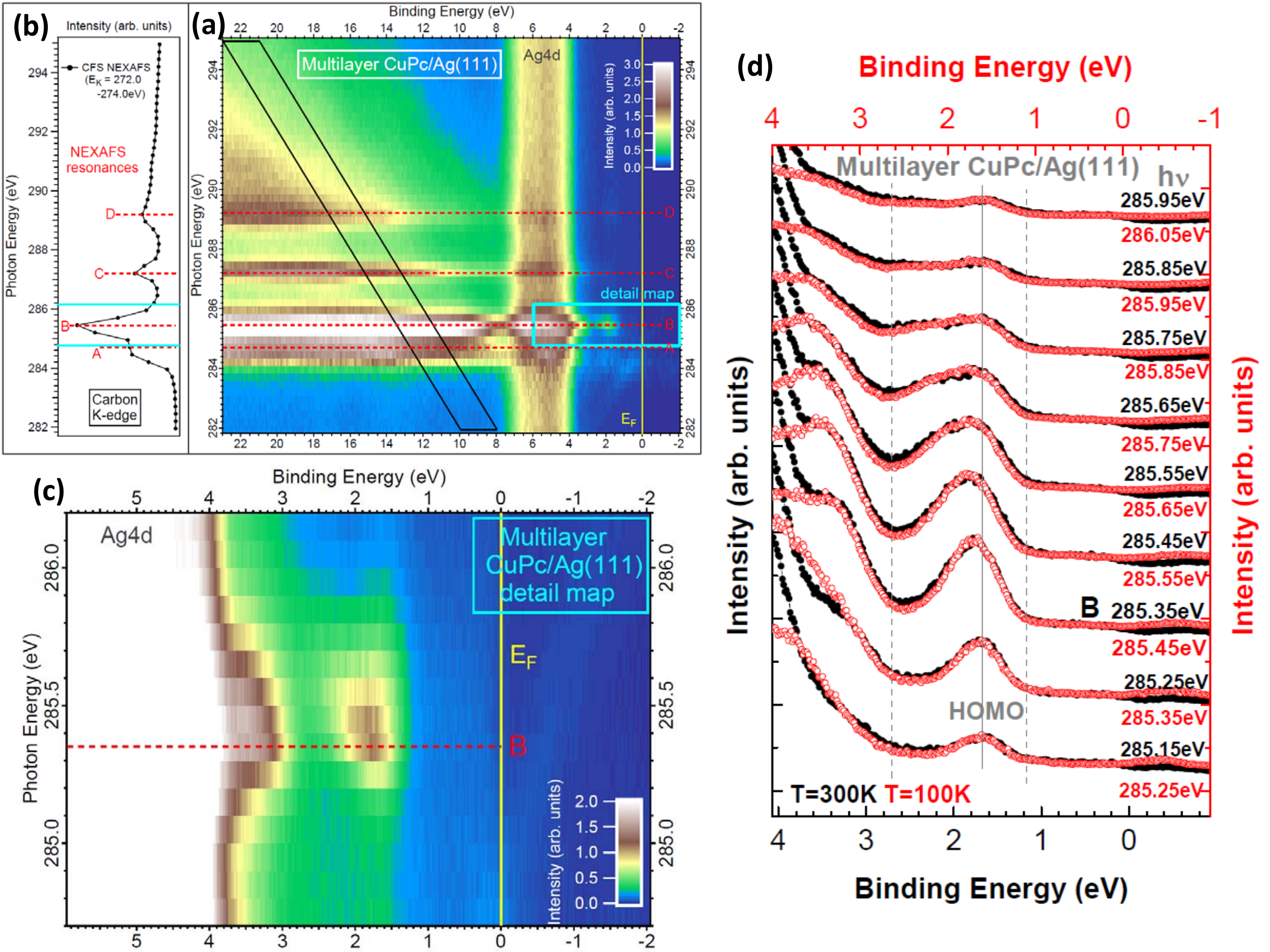}
    \caption{(color online). Full \mbox{RPES} data set of a multilayer CuPc/Ag(111) sample recorded at the carbon K-edge. The \mbox{PES} maps are normalized according to the procedure given in the text. \textbf{(a)} PES overview map with $h \nu$ increment of 250\,meV and $E_{B}$ increment of 50\,meV. \textbf{(b)} CFS \mbox{NEXAFS} extracted from the black box in the PES overview map (for details see text). The cyan lines denote the region probed by the PES detail map. \textbf{(c)} PES detail map with $h \nu$ increment of 100\,meV and $E_{B}$ increment of 15\,meV. The feature around 0\,eV $E_{B}$ at the bottom of the PES map that disperses to lower $E_{B}$ with increasing $h \nu$ is the the 2nd order C1s signal. \textbf{(d)} EDCs from the PES detail map (black) presented in a waterfall plot in comparison to corresponding EDCs from a PES detail map of a multilayer CuPc/Ag(111) sample prepared and measured at $T = 100$\,K (red). The EDCs of equivalent $h \nu$ positions within resonance B (resonance maximum denoted by the larger $h \nu$) are compared to each other (for details see text). The black axes ($E_{B}$ and intensity) correspond to the black EDCs and the red axes to the red EDCs. The red $E_{B}$ axis is shifted by -70\,meV with respect to the black one. The intensities of the $T = 300$\,K EDCs are unchanged while for each of the $T = 100$\,K EDCs an offset was added for a matching pre-peak plateau (differently damped substrate signal) and a scaling factor for best comparability of the line-shape of the HOMO signal is applied.}
    \label{fig:MultiCuPc}
\end{figure*}

\section{Experimental} \label{sec:Experimental}
\mbox{RPES} measurements were performed at BESSY II at the undulator beamline UE52-PGM ($E / \Delta E > 14000$ at 400\,eV photon energy, with cff=10 and 20\,$\mu$m exit slit \cite{RoccoJChemPhys}) in a UHV chamber with a pressure below \mbox{$5\cdot10^{-10}$\,mbar}. All PES maps were recorded with \mbox{p-polarized} light in 70° angle of incidence with respect to the surface normal, a beamline exit slit of 40\,$\mu$m and a cff value of 10. This results in an $h\nu$ resolution better than 40\,meV at 290\,eV. Photoelectron intensities were measured with a Scienta R4000 electron analyzer which was operated in transmission mode with an entrance slit of 300\,$\mu$m. For the PES overview maps (see Fig.\,\ref{fig:MultiCuPc}(a) and \ref{fig:MultiCoronene}(a)) a pass energy of 100\,eV and for the PES detail maps (see Fig.\,\ref{fig:MultiCuPc}(c), Fig.\,\ref{fig:MultiCoronene}(c) and all 1D energy distribution curve (EDC) waterfall plots) a pass energy of 50\,eV was used resulting in an energy resolution óf $\Delta E=150$\,meV for the former and $\Delta E=75$\,meV for the latter. $h \nu$ was calibrated with the Fermi edge ($E_F$) after the measurement of a PES map and the binding energy ($E_{B} = h\nu - E_{K}$) and $h \nu$ of the EDCs were shifted rigidly by the obtained offset. The resulting accuracy of both energy scales is better than $\Delta h \nu= \Delta E_{B} = 50$\,meV. Intensities were normalized to the ring current and the beamline flux curve which was recorded separately by measuring the $h \nu$ dependence of the intensity of a PES signal of the clean surface. 

The Ag(111) and Au(111) substrates were cleaned by several sputter and annealing cycles and their cleanness was confirmed by PES. All molecules were previously purified by sublimation and evaporated from Knudsen cells at pressures below $10^{-8}$\,mbar and with the substrate kept at 300\,K (unless denoted differently). The hetero-organic film was prepared by first depositing a multilayer of 3,4,9,10-perylene-tetracarboxylic-dianhydride (PTCDA), secondly tempering the film at 300°C which results in a single monolayer (ML) of PTCDA and finally depositing a submonolayer of CuPc. After each preparation step the sample was checked by PES measurements. Film thicknesses were determined by core level intensities of the uppermost layers, damping factors of underlying core level signals using the effective electron attenuation lengths given in Ref.\,\onlinecite{GraberSurfSci} and the emergence of multilayer originated features. The samples were carefully checked for radiation damage after the \mbox{RPES} data acquisition.

\section{Results and discussion}
In order to identify proper molecular orbital signals for a $h \nu$ dependent line-shape investigation our first step is to record a \mbox{PES} overview map (see Fig.\,\ref{fig:MultiCuPc}(a) and Fig.\,\ref{fig:MultiCoronene}(a)) with the $E_{B}$ ranging from approximately -2\,eV to 23\,eV and $h \nu$ including the intense \mbox{$\pi$-resonances}. In the resulting PES overview map we are then able to select the area in which the PES detail map (see Fig.\,\ref{fig:MultiCuPc}(c) and Fig.\,\ref{fig:MultiCoronene}(c)) is recorded in a subsequent measurement with smaller increments (see figure captions) for both $h \nu$ and $E_{B}$. While the $h \nu$ dependent $E_{B}$ dispersion is more obvious in the 2D PES detail map, a line-shape evolution can be observed better in a waterfall plot of 1D EDCs, especially when signals vary strongly in intensity. Thus for the general discussion of the $h \nu$ dependent line-shape evolution (Fig.\,\ref{fig:MultiCuPc} and Fig.\,\ref{fig:MultiCoronene}) we present the detail \mbox{RPES} data in both ways and restrict the data presentation to EDC waterfall plots for the following comparison of different systems (Fig.\,\ref{fig:MultiSnPcVsMultiCuPc} - Fig.\,\ref{fig:CoroneneSubMLVsCoroneneMulti}).

\subsection{The photoelectron spectroscopy map}
\begin{figure*}
    \centering
        \includegraphics[width=0.9\textwidth]{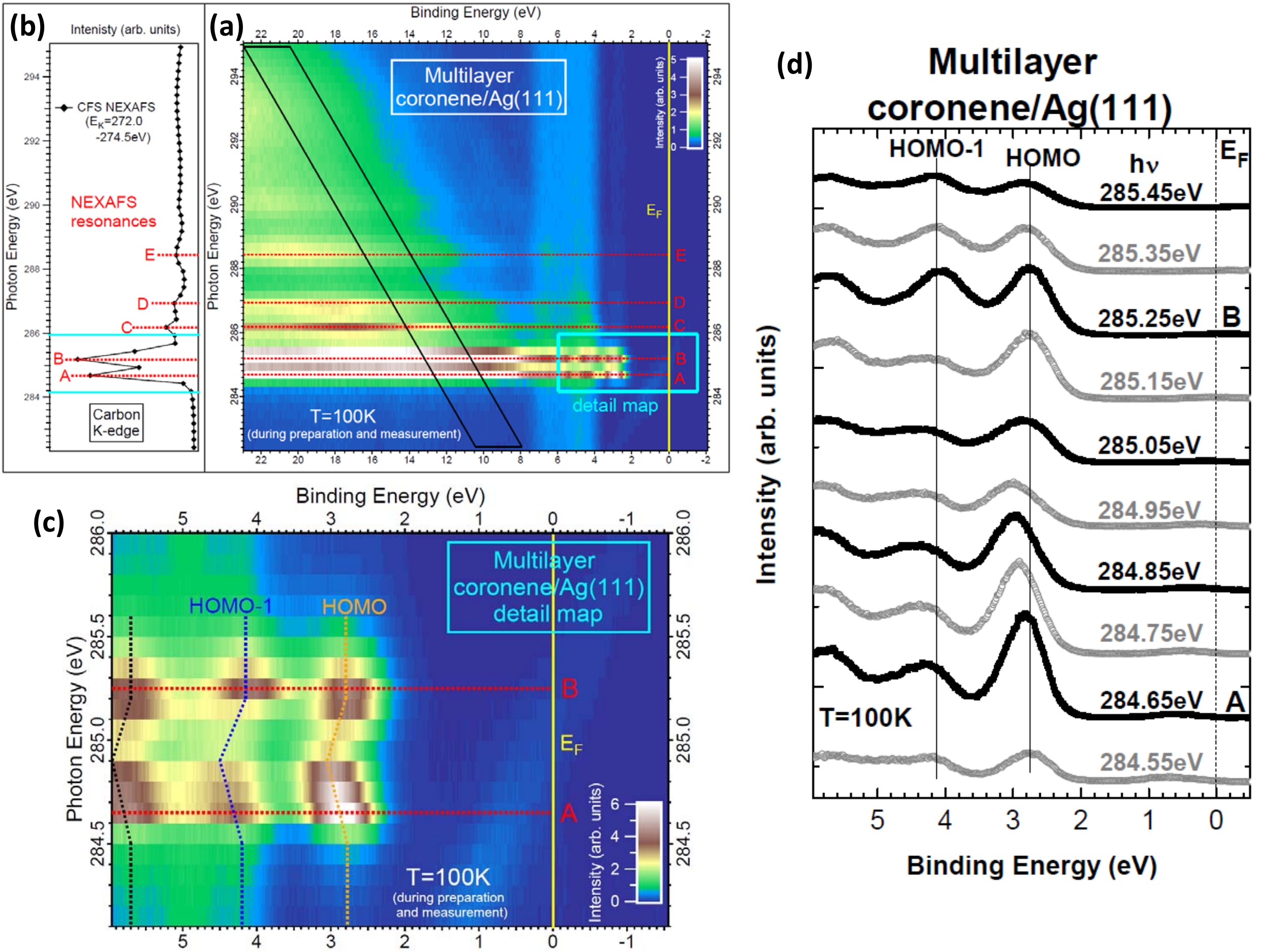}
    \caption{(color online). Full carbon K-edge \mbox{RPES} data set of a multilayer coronene/Ag(111) sample prepared and measured at $T = 100$\,K. The \mbox{PES} maps are normalized according to the procedure given in the text. \textbf{(a)} PES overview map with $h \nu$ increment of 250\,meV and $E_{B}$ increment of 50\,meV. \textbf{(b)} CFS \mbox{NEXAFS} extracted from the black box in the PES overview map (for details see text). The cyan lines denote the region of the PES detail map. \textbf{(c)} PES detail map with $h \nu$ increment of 100\,meV and $E_{B}$ increment of 15\,meV. The dotted vertical lines are a guide to the eye and represent the roughly parallel evolution of the MO peak maxima. The feature around 1\,eV $E_{B}$ at the bottom of the PES map that disperses to lower $E_{B}$ with increasing $h \nu$ is the the 2nd order C1s signal. \textbf{(d)} EDCs from the PES detail map presented in a waterfall plot.}
    \label{fig:MultiCoronene}
\end{figure*}

In the here presented study \mbox{RPES} is always performed at the carbon K-edge and the C1s electrons are excited into delocalized molecular orbitals with mainly \mbox{$\pi$-character}. On resonance PES signals stemming from the adsorbed organic molecules are massively enhanced with respect to the substrate PES signals. The states produced in the \mbox{RPES} process are one-hole final states along with additional two-hole final states which can further be divided into those with an additional electron in an excited state and those without. Above ionization threshold this electron cannot be present since it has been excited into vacuum. In the PES overview map (Fig.\,\ref{fig:MultiCuPc}(a) and Fig.\,\ref{fig:MultiCoronene}(a)) the features at high $E_{B}$ (8--23\,eV) consist of both two-hole final states and one-hole final states from lower lying molecular orbitals. These are located at the same $E_{B}$ as in an off-resonant (or direct) PES measurement. Integrating a constant $E_{K}$ window (black box in Fig.\,\ref{fig:MultiCuPc}(a) and Fig.\,\ref{fig:MultiCoronene}(a)) in this energetic region results in the 1D spectrum presented in Fig.\,\ref{fig:MultiCuPc}(b) and Fig.\,\ref{fig:MultiCoronene}(b). This approach is called constant final state (CFS) spectroscopy while the integration over a constant $E_{B}$ window is usually referred to as constant initial state (CIS) spectroscopy. For the obtained $h \nu$ increment the CFS spectrum is equal to the particular traditional partial electron yield \mbox{NEXAFS} spectrum while the CIS spectrum shows differences especially for the larger $h \nu$ region. This can be explained by the fact that above ionization threshold this region mainly consists of Auger signals which remain at constant $E_{K}$ with varying $h \nu$. The equivalence of this CFS NEXAFS spectrum and the partial electron yield \mbox{NEXAFS} means that for the assignment of the \mbox{NEXAFS} resonances the PES overview map itself can be used. This has the advantage that a comparison of intensities to another signal in the PES map is independent of the normalization procedure which might lead to delicate artifacts \cite{SchoelJourElSpec}.

The signals of interest in this work are the one-hole final states that are situated at lower $E_{B}$ (0--6\,eV) in the PES overview map. In the investigated large \mbox{$\pi$-conjugated} molecules these states stemming from the frontier molecular orbitals (HOMO etc.) are usually separated from each other in $E_{B}$. If the coupling to the surrounding is small they are also separated from all two-hole final state features. In this case a final state with two holes in the HOMO is the lowest bound state possible and this state is situated at considerably larger $E_{B}$. Thus they are ideal candidates for an electron-vibration coupling investigation by a $h \nu$ dependent line-shape analysis. Unfortunately the situation is expected to be more complicated in the $h \nu$ direction. A variety of close lying electronic transitions could contribute to the $h \nu$ region in which the separated signal from the molecular orbitals is intense enough for a significant line-shape analysis. Additionally multiple vibrational modes could simultaneously couple to each of these electronic transitions and to each other leading to an overlap of many contributions within a small $h \nu$ range. It turns out that this complicated situation can be overcome by a comparison to much simpler molecules. On the basis of this comparison to systems for which calculations are feasible the large \mbox{RPES} data set analyzed in this work can be discussed by including the character of the involved potential energy surfaces as the only parameters.

\subsection{Photon energy dependent line-shape evolution of molecular orbitals}
In Fig.\,\ref{fig:MultiCuPc} the complete \mbox{RPES} data set of a multilayer CuPc/Ag(111) is presented. Fig.\,\ref{fig:MultiCuPc}(a) shows the PES overview map from which the CFS \mbox{NEXAFS} was extracted (integration of black box in Fig.\,\ref{fig:MultiCuPc}(a) leads to the spectrum in Fig.\,\ref{fig:MultiCuPc}(b)) and used for the identification of the \mbox{NEXAFS} resonances A-D. A pronounced increase in intensity of the HOMO signal is observed in resonance B so this part of the PES overview map (cyan box) is selected for closer inspection by a PES detail map (Fig.\,\ref{fig:MultiCuPc}(c)). Here the resonance maximum (dashed red line) is defined to be the EDC with the largest intensity at the $E_{B}$ position of the HOMO signal in direct PES. Starting closely before the resonance maximum the maximum of the HOMO signal continuously disperses to higher $E_{B}$ with increasing $h \nu$. The EDC waterfall plot (Fig.\,\ref{fig:MultiCuPc}(d)) allows to identify this effect as a gradual redistribution of intensity towards higher $E_{B}$ within the HOMO signal. This results in a line-shape evolution from symmetric to more and more asymmetric and finally to two separated peaks. The black EDCs which originate from the PES detail map in Fig.\,\ref{fig:MultiCuPc}(c) hereby evolve parallel to the red EDCs which stem from a corresponding PES detail map of a multilayer CuPc prepared and measured at $T = 100$\,K. The EDCs are aligned at the same position within resonance B, i.e. both EDCs of the resonance maximum are compared to each other and so on. Interestingly the EDCs of the 300\,K sample, which grows in a Stranski-Krastanow mode, have to be shifted in $h \nu$ and $E_{B}$ for comparison to the 100\,K sample exhibiting layer-by-layer type growth. However, both samples show an identical line-shape evolution. An imprecise energy calibration can be excluded as a reason for the energy mismatch since the HOMO $E_{B}$ positions exhibit the same shift in the corresponding off-resonant ($h\nu$=120\,eV) PES data. Apparently the difference in growth mode influences the peak positions in (R)PES and \mbox{NEXAFS} spectroscopy but not the origin of the $h \nu$ dependent line-shape evolution. This is further confirmed by a comparison to the EDCs of a multilayer CuPc/Au(111) which exhibit the same line-shape evolution as the two systems displayed in Fig.\,\ref{fig:MultiCuPc}(d).

\begin{figure}
    \centering
        \includegraphics[width=0.45\textwidth]{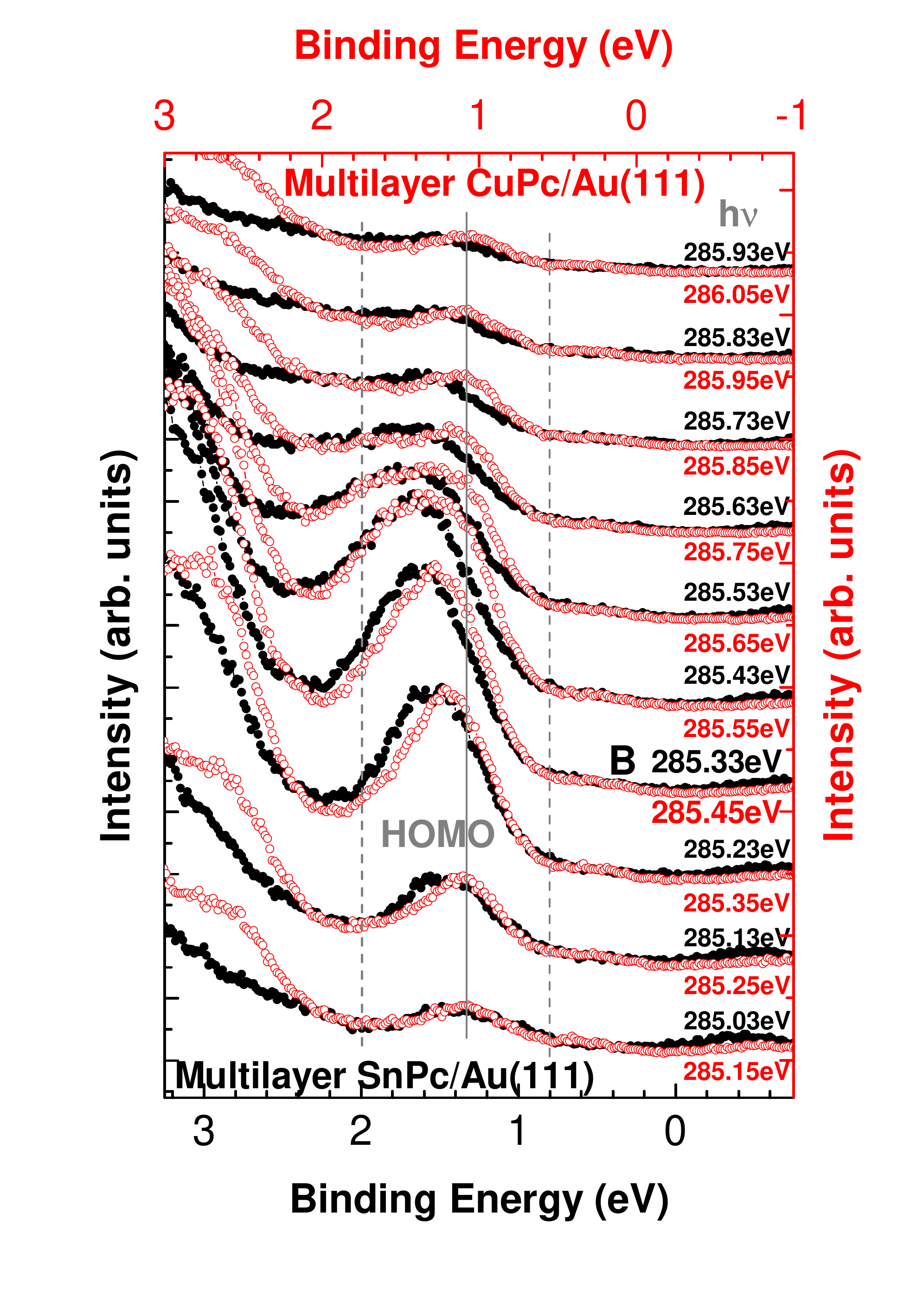}
    \caption{(color online). Waterfall plot of EDCs from a multilayer CuPc/Au(111) (red) compared to EDCs from a multilayer of SnPc/Au(111) (black). The EDCs of equivalent $h \nu$ positions within the resonance (resonance maximum denoted by the larger $h \nu$) are compared to each other (for details see text). The black axes ($E_{B}$ and intensity) correspond to the black EDCs and the red axes to the red EDCs. The red $E_{B}$ axis was shifted by -250\,meV with respect to the black one. The detail PES maps from which the EDCs originate were normalized according to the procedure given in the text. The intensities of the SnPc EDCs are unchanged while each of the CuPc EDCs is scaled for best comparability of the line-shape of the HOMO signal.}
    \label{fig:MultiSnPcVsMultiCuPc}
\end{figure}

Up to now a proper treatment of electronic interference effects in the \mbox{RPES} process \cite{FanoPhysRev,DavisPRB} or lifetime vibrational interference \cite{NeebJourElSpec,OsborneJChemPhys} is not feasible due to the complexity of the electronic and vibrational structure of the investigated large \mbox{$\pi$-conjugated} molecules. Not even the assignment of electronic transitions that contribute to the \mbox{NEXAFS} resonance B of CuPc is unambiguously determined in literature \cite{EvangelistaJChemPhys,LinaresJourPhysChemB,FrancescoJPhysChemA}. Hence we cannot draw any conclusion from theoretical calculations performed on the here investigated system. Instead we compare the \mbox{RPES} data of CuPc to studies of smaller molecules for which \mbox{RPES} calculations are feasible. In several studies of small molecules \cite{OsborneJChemPhys,PiancastelliJPhysB,NeebJourElSpec,TravnikovaJourElSpec} (CO, N$_2$, O$_2$ and OCS) a quite similar $h \nu$ dependent line-shape evolution of the molecular orbital signals (usually denoted as participator decays in these publications) is observed. The case of CO constitutes the simplest scenario of one electronic transition in the \mbox{NEXAFS} spectrum and one vibration coupling to it. Therefore the different vibronic peaks contributing to the molecular orbital signals can be resolved and a selective excitation into one particular vibrational level ($\nu_i=0,1,2$) of the intermediate state potential energy surface ($V_i$) in \mbox{RPES} is possible. In CO it is found that the higher the $\nu_i$ of the excitation into the $V_i$ is the more relative intensity is distributed into higher vibrational final state levels $\nu_f$ of the molecular orbital signals. In a Franck-Condon picture this means that the character of the potential energy surfaces govern the relative intensities of the different $\nu_f$ contributions to the molecular orbital signal for a particular $h \nu$. Relative intensity changes between the $\nu_f$ peaks for EDCs recorded with different $h \nu$ are thus the reason for the line-shape evolution. If the different $\nu_f$ peaks cannot be resolved this electron-vibration coupling originated effect leads to successively growing relative intensity in the high $E_{B}$ part of the molecular orbital peak with increasing $h \nu$. This is exactly the observation seen for the CuPc HOMO signal in Fig.\,\ref{fig:MultiCuPc}(d). So we conclude that this $h \nu$ dependent line-shape evolution can be discussed analogously to CO. Hence the excitation into different $\nu_i$ and the consequential characteristic distribution of relative intensity (Franck-Condon-factors) of the $\nu_f$ is the basis for the following discussion. Consequently the character of the potential energy surface in the ground state ($V_g$), in the intermediate state ($V_i$) and in the final state ($V_f$) are the parameters which determine the line-shape evolution in the picture we use to discuss the large \mbox{$\pi$-conjugated} molecules in this work. Furthermore the observation of a continuous line-shape evolution in Fig.\,\ref{fig:MultiCuPc}(d) points towards one dominating electronic contribution to the \mbox{RPES} in $h \nu$ direction so that despite its complicated electronic structure CuPc can be compared to CO. Multiple electronic transitions with a small difference in $h \nu$ would also result in a continuous line-shape evolution and cannot be excluded. In this case we would have to speak of one effective electronic transition that is responsible for the observed HOMO signal in \mbox{RPES} in Fig.\,\ref{fig:MultiCuPc}(d). However, the comparability to CO is reasonable in both cases.

We would like to note that in a previous discussion of the \mbox{RPES} data of SnPc the line-shape evolution of molecular orbital signals was explained by $h \nu$ detuning from the contributing electronic transition in the \mbox{NEXAFS} spectrum \cite{HaemingChemPhysLett}. Based on the present large data set we consider the above described excitation into different $\nu_i$ the dominating effect for the observed line-shape changes within the investigated $h \nu$ regime.

For a multilayer coronene/Ag(111) a complete \mbox{RPES} data set is presented in Fig.\,\ref{fig:MultiCoronene} analogous to CuPc in Fig.\,\ref{fig:MultiCuPc}. The PES detail map (Fig.\,\ref{fig:MultiCoronene}(c)) and the EDC waterfall plot (Fig.\,\ref{fig:MultiCoronene}(d)) of this system do not show the same overall continuous line-shape evolution as for CuPc. In coronene the HOMO peak maximum first disperses to higher $E_{B}$ with increasing $h \nu$ starting at the maximum of resonance A. Then, between the resonances A and B, it moves back to its original $E_{B}$ position and remains almost unchanged with further increasing $h \nu$. This non-continuous evolution scenario is quite similar to H$_2$CO \cite{BozekChemPhys} and NO \cite{KukkPRA}. In the former multiple simultaneously exited vibronic modes contribute to the \mbox{NEXAFS} spectrum with one electronic transition \cite{RemmersPRA}. In the latter multiple electronic transitions are present in the \mbox{NEXAFS} spectrum with the single vibrational mode of NO coupling to them \cite{KukkPRA}. Both scenarios are generally possible for the $h \nu$ region from 284.4\,eV to 285.6\,eV in the \mbox{RPES} data of coronene. \mbox{NEXAFS} calculations suggest that there is one electronic transition in resonance A and another two in resonance B \cite{OjiJChemPhys} and the line-shape of the high resolution partial electron yield \mbox{NEXAFS} spectrum (not shown) corroborates this scenario of multiple electronic transitions. A significant contribution to the \mbox{RPES} from the single transition in resonance A is clearly observed. Hence for the $h \nu$ region around resonance A ($h \nu = 284.55$\,eV$ - 284.95$\,eV) the HOMO signal line-shape evolution of coronene can be explained in the same way as for CuPc described above. Further data on a sample of a multilayer coronene/Ag(111) prepared and measured at 300\,K and recorded with a 50\,meV $h \nu$ increment (not shown) confirms the continuous evolution observed for the HOMO signal around resonance A. Whether there is a contribution to the \mbox{RPES} from none, one or two electronic transitions in \mbox{NEXAFS} resonance B cannot be concluded from the data. However in any of these cases the intensity and line-shape evolution can be discussed analogously since for a second electronic transition as well as a higher excited vibrational series of the first electronic transition higher $\nu_i$ are excited by higher $h \nu$. The $E_{B}$ dispersion of the HOMO peak in between resonance A and B is most probably a consequence of the decreasing intensity at higher $\nu_i$ within \mbox{NEXAFS} resonance A and a simultaneous increase of the intensity at low $\nu_i$ within \mbox{NEXAFS} resonance B. So a peak with more relative intensity at higher $E_{B}$ gradually transforms into a peak with more relative intensity at lower $E_{B}$. For the $h \nu$ region around resonance B ($h \nu = 285.15$\,eV $- 285.45$\,eV) a less pronounced redistribution of relative intensity to higher $E_{B}$ with increasing $h \nu$ is found. In the Franck-Condon picture discussed above this is a consequence of a difference of the $V_i$. The fact that all three visible molecular orbital signals in Fig.\,\ref{fig:MultiCoronene}(c) evolve in an almost parallel way (see dotted vertical lines which roughly represent the peak maximum) but different in resonances A and B further corroborates this conclusion. The observed resonance specific line-shape evolution requires an explanation with a parameter determined by the excitation such as $V_i$. The parallel evolution then suggests similar $V_f$ for the three molecular orbital signals which is reasonable due to the \mbox{$\pi$-dominated} character of these frontier molecular orbitals.

\begin{figure}
    \centering
        \includegraphics[width=0.45\textwidth]{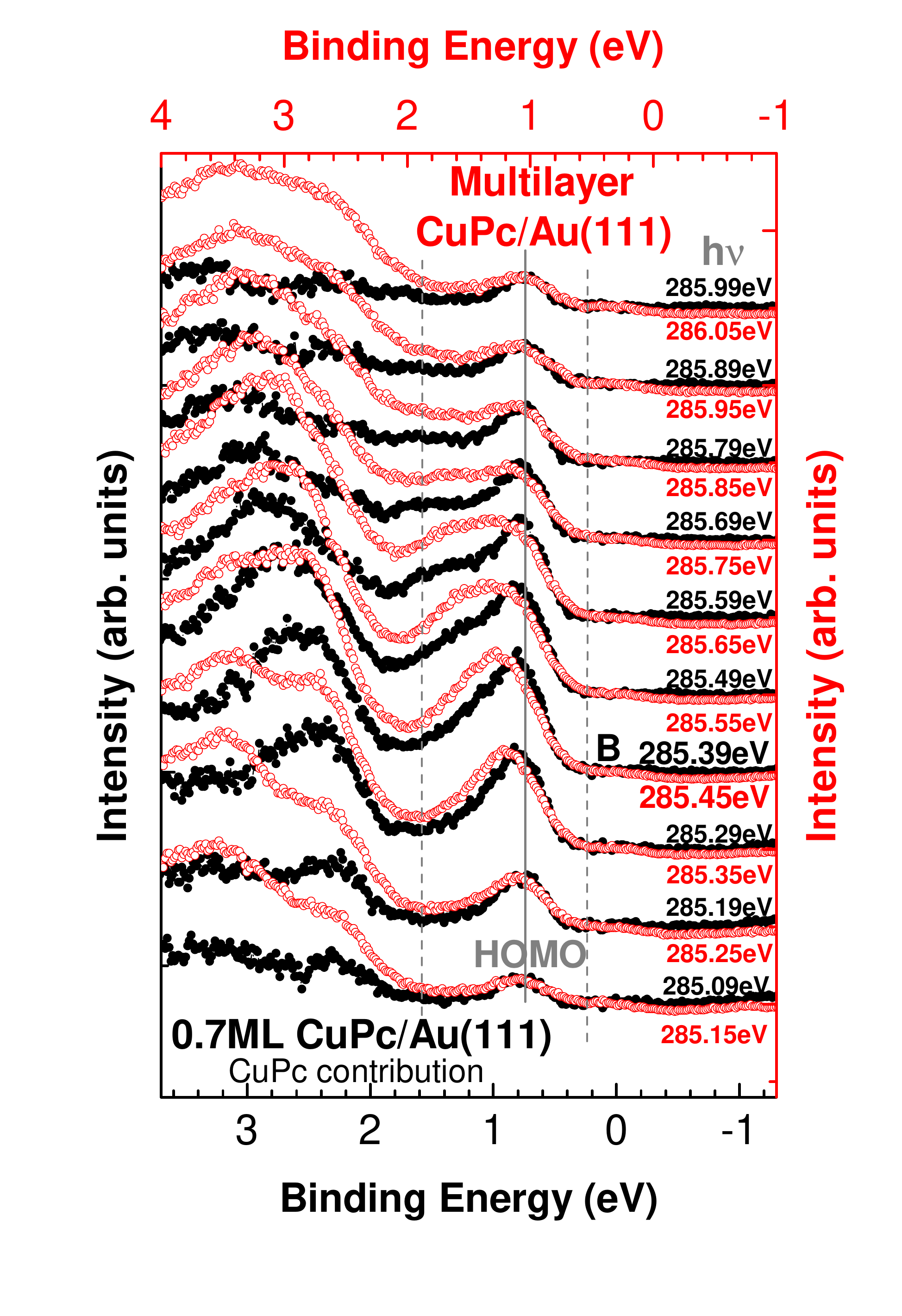}
    \caption{(color online). Waterfall plot of EDCs from a multilayer CuPc/Au(111) (red) compared to EDCs from a 0.7\,ML CuPc/Au(111) (black). The EDCs of equivalent $h \nu$ positions within the resonance (resonance maximum denoted by the larger $h \nu$) are compared to each other (for details see text). The black axes ($E_{B}$ and intensity) correspond to the black EDCs and the red axes to the red EDCs. The red $E_{B}$ axis is shifted by 300\,meV with respect to the black one. The PES detail maps from which the EDCs originate are normalized according to the procedure given in the text. For the 0.7\,ML CuPc/Au(111) EDCs the scaled Au(111) substrate EDCs at equal $h \nu$ are subtracted and the intensities are unchanged. To each of the multilayer CuPc EDCs an offset is added for a matching pre-peak plateau (clean Au(111) subtracted only for the 0.7\,ML CuPc EDCs) and a scaling factor for best comparability of the line-shape of the HOMO signal is applied.}
    \label{fig:CuPcSubMLAuVsCuPcMulti}
\end{figure}

The comparison of the $h \nu$ dependent line-shape evolution in \mbox{RPES} for CuPc and SnPc multilayers displayed in Fig.\,\ref{fig:MultiSnPcVsMultiCuPc} serves as another consistency check for the consideration of the character of the potential energy surfaces as the dominant parameters. Both molecules have a very similar overall electronic structure and in particular a basically equal HOMO because there is no contribution by the center metal atom \cite{AristovJChemPhys,StadlerNaPhys} to it. Thus the geometric difference between the two molecules constitutes the most reasonable explanation for the observed difference visible in the HOMO signal line-shape evolution. A difference in electron-vibration coupling in the \mbox{RPES} process is not surprising for the bent SnPc with respect to the flat CuPc. Hence the character of the potential energy surfaces can also explain the origin of the difference between these two systems. The fact that both HOMO signal line-shapes are almost equal at the lowest $h \nu$ in Fig.\,\ref{fig:MultiSnPcVsMultiCuPc} but then evolve differently throughout the resonance reveals the sensitivity of the technique. Even in a moderate $E_{B}$ resolution leading to symmetric and similar peaks for both molecules in direct PES a difference in the HOMO signal line-shape evolution is found. In the following section we explore the potential of this line-shape evolution in \mbox{RPES} as a probe for variations in the molecular environment.

\subsection{Modification of the electron-vibration coupling by different interfaces}

\begin{figure}
    \centering
        \includegraphics[width=0.45\textwidth]{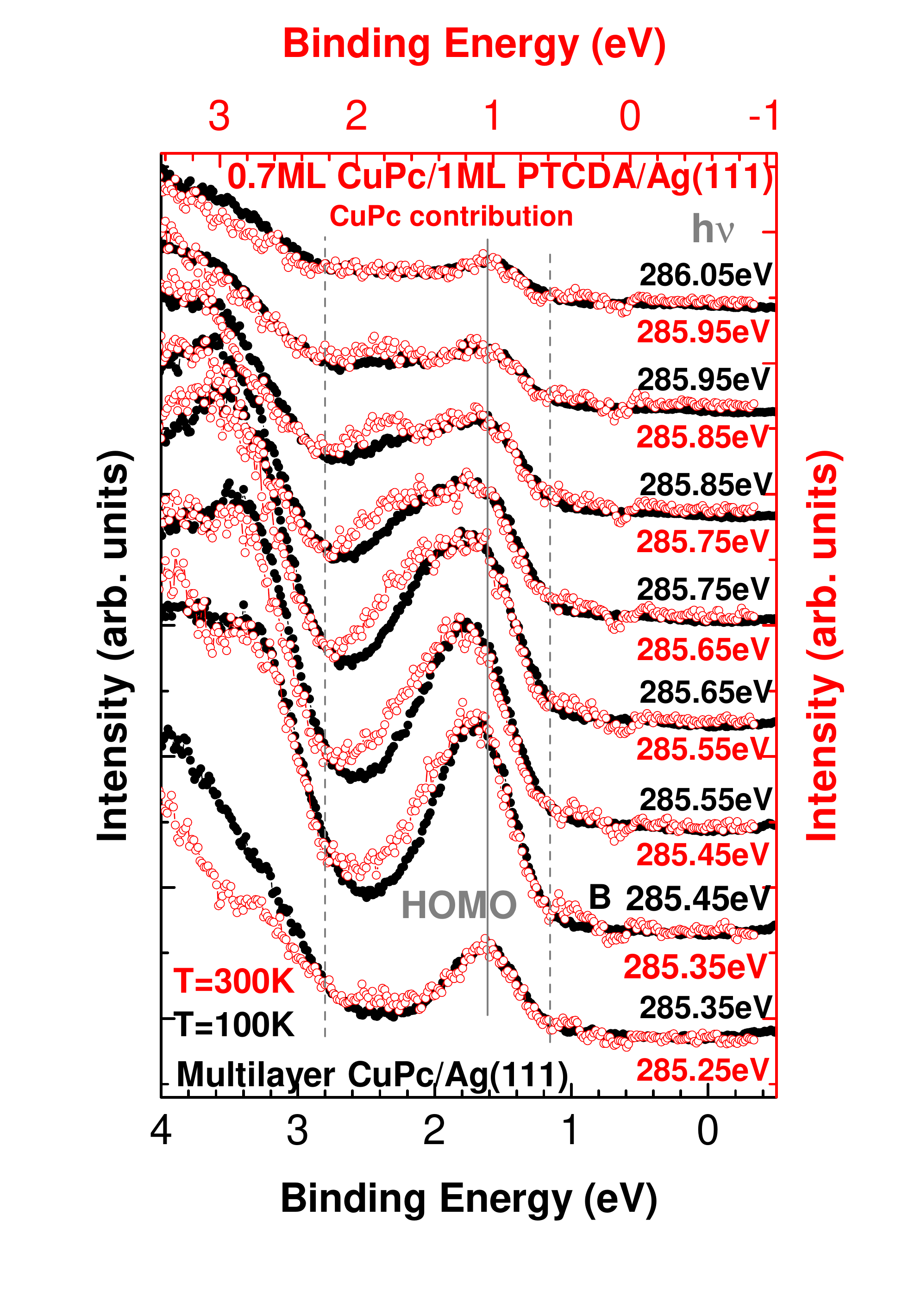}
    \caption{(color online). Waterfall plot of EDCs from a multilayer CuPc/Ag(111) prepared and measured at $T = 100$\,K (black) compared to EDCs from a 0.7\,ML CuPc/1\,ML PTCDA/Ag(111) (red). The EDCs of equivalent $h \nu$ positions within the resonance (resonance maximum denoted by the larger $h \nu$) are compared to each other (for details see text). The black axes ($E_{B}$ and intensity) correspond to the black EDCs and the red axes to the red EDCs. The red $E_{B}$ axis is shifted by -570\,meV with respect to the black one. The PES detail maps from which the EDCs originate are normalized according to the procedure given in the text. For the 0.7\,ML CuPc/1\,ML PTCDA/Ag(111) EDCs scaled 1\,ML PTCDA/Ag(111) EDCs at equal $h \nu$ (from Ref.\,\onlinecite{ZiroffUnpub}) are subtracted and each EDC is scaled for best comparability of the line-shape of the HOMO signal.}
    \label{fig:CuPcSubMLPTCDAVsMulti}
\end{figure}

\begin{figure*}
    \centering
        \includegraphics[width=0.9\textwidth]{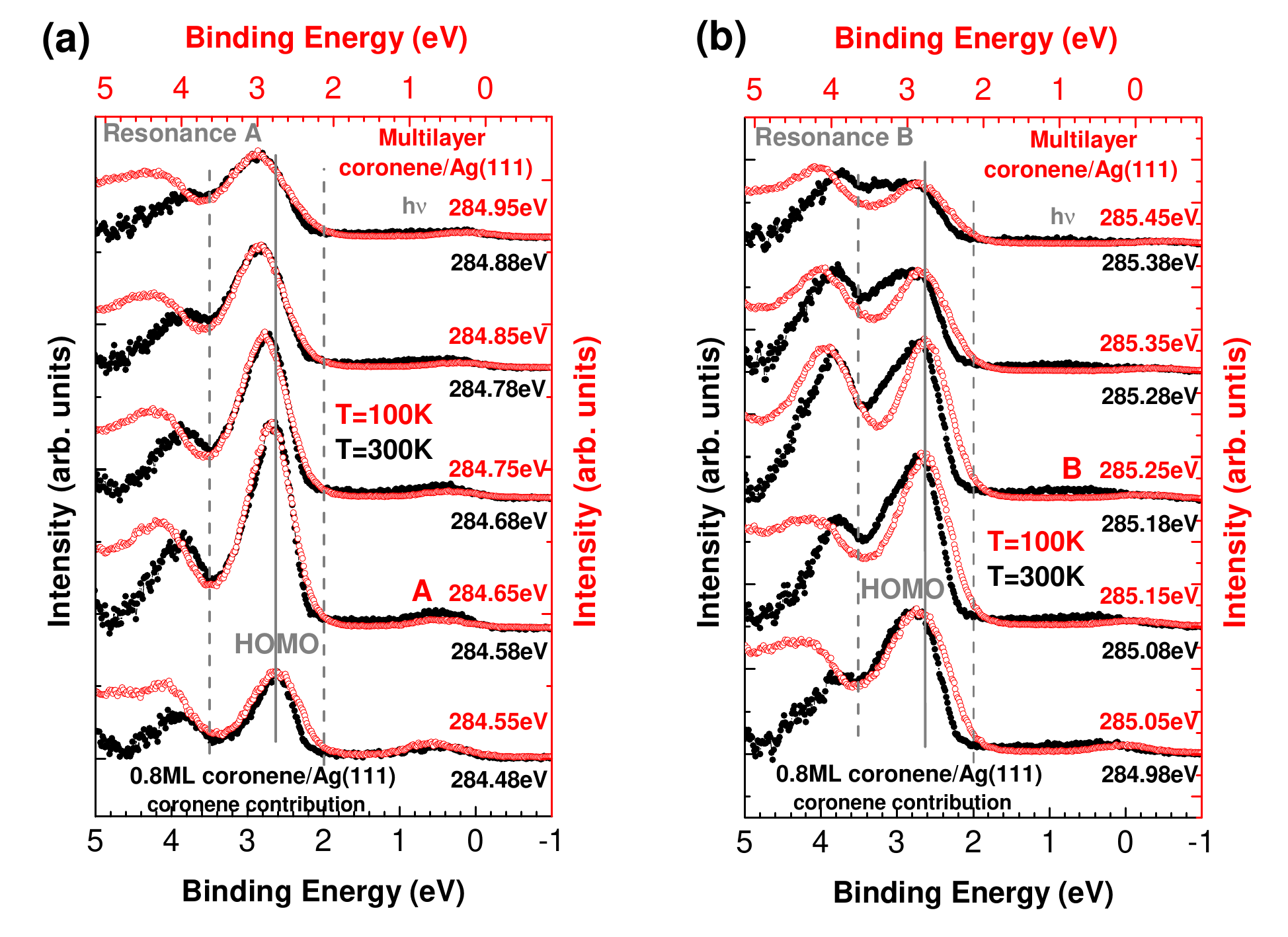}
    \caption{(color online). Waterfall plot of EDCs from a multilayer coronene/Ag(111) prepared and measured at $T = 100$\,K (red) compared to EDCs from a 0.8\,ML coronene/Ag(111) (black). The EDCs of equivalent $h \nu$ positions within the resonance (resonance maximum denoted by the capital letters A and B) are compared to each other (for details see text). The black axes ($E_{B}$ and intensity) correspond to the black EDCs and the red axes to the red EDCs. The red $E_{B}$ axis is shifted by 130\,meV with respect to the black one. The PES detail maps from which the EDCs originate are normalized according to the procedure given in the text. For the 0.8\,ML coronene/Ag(111) EDCs scaled Ag(111) substrate EDCs at equal $h \nu$ are subtracted and the intensities are unchanged. Each of the multilayer coronene EDCs is scaled for best comparability of the line-shape of the HOMO signal. \textbf{(a)} $h \nu$ region of \mbox{NEXAFS} resonance A. \textbf{(b)} $h \nu$ region of \mbox{NEXAFS} resonance B.}
    \label{fig:CoroneneSubMLVsCoroneneMulti}
\end{figure*}

Comparing the line-shape evolution of a multilayer CuPc/Au(111) with a submonolayer CuPc/Au(111) in Fig.\,\ref{fig:CuPcSubMLAuVsCuPcMulti} exposes a clear difference between the two systems. As already seen in Fig.\,\ref{fig:MultiSnPcVsMultiCuPc} an equal line-shape at the lowest $h \nu$ evolves differently at higher $h \nu$. In this example the difference can be characterized as less distribution of relative intensity into the higher $E_{B}$ part of the HOMO signal for CuPc adsorbed on Au(111). In other words the excitation of higher $\nu_f$ in the \mbox{RPES} process is somewhat prohibited. Such an influence due to the adsorption on a metal surface on the character of the involved potential energy surfaces is reasonable and thus fits into the overall discussion. A comparison of electron-vibration coupling in direct PES of submonolayer or ML samples with respect to multilayer films is usually hindered by broadening due to disorder in the multilayer. In \mbox{RPES} the involvement of an additional $V_i$ and the possibility to excite high $\nu_i$ in this $V_i$ results in changes of the line-shape beyond the broadening due to disorder. So for the here investigated molecules the $h \nu$ dependent line-shape evolution of molecular orbital signals in \mbox{RPES} allows to identify interface originated modifications of electron-vibration coupling. Additionally both systems compared in Fig.\,\ref{fig:CuPcSubMLAuVsCuPcMulti} are usually characterized as weakly interacting because there is no significant electronic interaction that manifests itself in strongly altered PES spectra of the adsorbed molecule with respect to the multilayer film. The revelation of differences within this so called physisorptive regime further indicates the sensitivity of the effect. Remarkably the analogous comparison of a submonolayer SnPc/Au(111) with a multilayer SnPc/Au(111) (not shown) exhibits a parallel $h \nu$ dependent line-shape evolution of the SnPc HOMO signal. Hence the potential energy surfaces of SnPc are not significantly altered by the adsorption on Au(111). Consequently the electron-vibration coupling of the bent SnPc molecule not only differs from the one of the flat CuPc but also a difference in the influence of the Au(111) surface on the electron-vibration coupling in both molecules is observed.

By depositing a submonolayer or ML of SnPc \cite{HaemingPRB,HaemingSurfSci} or CuPc \cite{StadtmullerPRL} onto 1\,ML PTCDA/Ag(111) a stable and vertically well defined interface can be prepared. The comparison of the $h \nu$ dependent line-shape evolution of the CuPc HOMO signal of this hetero-organic system with a multilayer of CuPc/Ag(111) is displayed in Fig.\,\ref{fig:CuPcSubMLPTCDAVsMulti}. In contrast to the adsorption on Au(111) the excitation of higher $\nu_f$ in the \mbox{RPES} process is enhanced at this interface with respect to the multilayer film. For the SnPc/1\,ML PTCDA/Ag(111) hetero-molecular interface the same variation is found with respect to the SnPc multilayer \cite{HaemingChemPhysLett}. The smaller $h \nu$ increment of the here presented data makes a clear identification of a continuous evolution possible and therefore shows that both SnPc and CuPc in the hetero-molecular system can be discussed in the same way as all other systems in this work. The similar change of both molecules in the hetero-molecular system with respect to a multilayer sample becomes particularly interesting considering the fact that the reaction on the adsorption on Au(111) is different. We would like to note that additional \mbox{RPES} data of several systems with strong interfacial bonding and ground state charge transfer into the molecule (1\,ML PTCDA/Ag(110), 1\,ML 1,4,5,8-naphthalene-tetracarboxylic dianhydride/Ag(111) and 1\,ML CuPc/Ag(111)) show very strong and characteristic line-shape changes with respect to the corresponding multilayer sample and have to be described within a different model. Hence in terms of interface modifications of electron-vibration coupling observed in \mbox{RPES} both SnPc and CuPc on PTCDA/Ag(111) have to be considered weakly interacting, which is in contrast to another conclusion in literature \cite{StadtmullerPRL}. A detailed investigation of \mbox{RPES} applied to the strongly interacting systems mentioned above will be given in a future publication.

For a multilayer of coronene a different line-shape evolution in \mbox{RPES} is found for the \mbox{NEXAFS} resonances A and B (see Fig.\,\ref{fig:MultiCoronene}). This discussed difference in the character of the particular $V_i$ also manifests itself in the reaction on the adsorption on a Ag(111) surface. A comparison of the submonolayer film on Ag(111) to the multilayer sample is displayed in Fig.\,\ref{fig:CoroneneSubMLVsCoroneneMulti}. The $h \nu$ region around resonance A (Fig.\,\ref{fig:CoroneneSubMLVsCoroneneMulti}(a)) shows a parallel line-shape evolution starting with the resonance maximum while for the EDCs at the lowest $h \nu$ a broader HOMO signal is observed for the multilayer. This difference can be explained by increased inhomogeneity in the multilayer film. However, this effect becomes negligible as soon as the electron-vibration coupling originated line-shape evolution alters the HOMO signal to a much larger extend. In the $h \nu$ region around resonance B (Fig.\,\ref{fig:CoroneneSubMLVsCoroneneMulti}(b)) the submonolayer sample EDCs exhibit more intensity in the high $E_{B}$ part of the HOMO peak than the multilayer EDCs. Thus the Ag(111) surface modifies the $V_i$ such that the excitation of higher $\nu_f$ in the \mbox{RPES} process is enhanced with respect to the multilayer. A comparison of the line-shape evolution of two coronene multilayers prepared and measured at 100\,K and at 300\,K (not shown) further corroborates the assignment of the difference in Fig.\,\ref{fig:CoroneneSubMLVsCoroneneMulti}(b) as an interface effect. The difference in growth modes (300\,K grows in a Stranski-Krastanow mode, 100\,K more in a layer-by-layer mode) has no significant influence on the line-shape evolution analogous to the finding in CuPc discussed above. Consequently the example of a submonolayer of coronene/Ag(111) shown in Fig.\,\ref{fig:CoroneneSubMLVsCoroneneMulti} reveals that the interface originated modification of the electron-vibration coupling in \mbox{RPES} can even be characteristic for different \mbox{NEXAFS} resonances of the same system. This once again points towards a change of the $V_i$ as the most reasonable origin for the observed differences in the $h \nu$ dependent line-shape evolution since these are resonance specific.

\section{Summary and conclusion}
In summary we present a comprehensive \mbox{RPES} study of several large \mbox{$\pi$-conjugated} molecules in multilayer films and at different interfaces. By the comparison to \mbox{RPES} investigations of small molecules we are able to discuss the observed $h \nu$ dependent line-shape evolution of molecular orbital signals within a picture based on electron-vibration coupling. The relative intensity distribution among excitations of vibrations in the final state for a particular $h \nu$ is suggested as the origin for the observed line-shape evolution. Within this simplified picture all systems and the differences in between them are consistently discussed by the character of the potential energy surfaces involved in the \mbox{RPES} process. Consequently the influence of the particular interface on the mechanism of the excitation of these vibrations is given as the origin for changes with respect to the multilayer sample. We show that depending on the specific interface the relative intensity distributed into high vibrational levels in the final state can remain unchanged, can be partially quenched or enhanced. On Au(111) the excitation of high vibrational levels is partially quenched for CuPc while no influence is found for SnPc. Opposite to that, an enhancement of the high vibrational levels occurs at the hetero-organic interface of SnPc and CuPc on top of \mbox{1\,ML PTCDA/Ag(111)}. With the adsorption of coronene on Ag(111) we present an example for a resonance specific modification of electron-vibration coupling. In comparison to the multilayer film an enhancement of the contribution of high vibrations to the HOMO signal is demonstrated only in one of the two \mbox{NEXAFS} resonances. 

This work illustrates the potential of line-shape investigations of molecular orbital signals in \mbox{RPES} as a sensitive probe for electron-vibration coupling in large \mbox{$\pi$-conjugated} molecules. By the presented magnitude of the $h \nu$ dependent variation inhomogeneous broadening and moderate energy resolution can be overcome which allows to reveal effects that are not visible in direct PES and \mbox{NEXAFS} spectroscopy alone. Therefore a smaller influence of an interface on electron-vibration coupling can be investigated by \mbox{RPES} with an equally high surface sensitivity. The fact that all the interface originated modifications found constitute differences within the physisorptive regime further demonstrates the sensitivity of the applied technique.

A detailed investigation of the specific character of the potential energy surfaces involved in \mbox{RPES}, like the role of coupling in between multiple vibronic modes and lifetime vibrational interference can only be studied by direct comparison of our data with calculations. These are to our best knowledge not feasible for the investigated systems at this point. The same applies for electronic interference effects or a possible insufficiency of the Born-Oppenheimer approximation. We would like to stress that the importance of these effects cannot be excluded on the basis of this experimental work. However, the given discussion based on the comparison to small molecules draws a consistent picture for a large data set using the character of the potential energy surfaces as the only required parameters. With this work we hope to stimulate a significant advance in the theoretical treatment of \mbox{RPES} applied to the here presented systems which is necessary to elevate the understanding of this technique to the next level.

\acknowledgments{We thank the BESSY staff, F. Bruckner, S. Gusenleitner and M. Scholz for support during beamtimes. Further we would like to thank F. Meyer for stimulating discussions about \mbox{RPES} in comparison to RIXS. This work was supported by BESSY, by the Bundesministerium f\"ur Bildung und Forschung BMBF (grant no. 05K10WW2 and 03SF0356B) and by the Deutsche Forschungsgemeinschaft DFG (GRK 1221 and RE1469/9-1).}


\begin{thebibliography}{10}%
\makeatletter
\providecommand \@ifxundefined [1]{%
 \ifx #1\undefined \expandafter \@firstoftwo
 \else \expandafter \@secondoftwo
\fi
}%
\providecommand \@ifnum [1]{%
 \ifnum #1\expandafter \@firstoftwo
 \else \expandafter \@secondoftwo
\fi
}%
\providecommand \enquote [1]{``#1''}%
\providecommand \bibnamefont  [1]{#1}%
\providecommand \bibfnamefont [1]{#1}%
\providecommand \citenamefont [1]{#1}%
\providecommand\href[0]{\@sanitize\@href}%
\providecommand\@href[1]{\endgroup\@@startlink{#1}\endgroup\@@href}%
\providecommand\@@href[1]{#1\@@endlink}%
\providecommand \@sanitize [0]{\begingroup\catcode`\&12\catcode`\#12\relax}%
\@ifxundefined \pdfoutput {\@firstoftwo}{%
 \@ifnum{\z@=\pdfoutput}{\@firstoftwo}{\@secondoftwo}%
}{%
 \providecommand\@@startlink[1]{\leavevmode}%
 \providecommand\@@endlink[0]{}%
}{%
 \providecommand\@@startlink[1]{%
  \leavevmode
  \pdfstartlink
   attr{/Border[0 0 1 ]/H/I/C[0 1 1]}%
   user{/Subtype/Link/A<</Type/Action/S/URI/URI(#1)>>}%
  \relax
 }%
 \providecommand\@@endlink[0]{\pdfendlink}%
}%
\providecommand \url  [0]{\begingroup\@sanitize \@url }%
\providecommand \@url [1]{\endgroup\@href {#1}{\urlprefix}}%
\providecommand \urlprefix [0]{URL }%
\providecommand \Eprint[0]{\href }%
\@ifxundefined \urlstyle {%
  \providecommand \doi [1]{doi:\discretionary{}{}{}#1}%
}{%
  \providecommand \doi [0]{doi:\discretionary{}{}{}\begingroup
  \urlstyle{rm}\Url }%
}%
\providecommand \doibase [0]{http://dx.doi.org/}%
\providecommand \Doi[1]{\href{\doibase#1}}%
\providecommand \bibAnnote [3]{%
  \BibitemShut{#1}%
  \begin{quotation}\noindent
    \textsc{Key:}\ #2\\\textsc{Annotation:}\ #3%
  \end{quotation}%
}%
\providecommand \bibAnnoteFile [2]{%
  \IfFileExists{#2}{\bibAnnote {#1} {#2} {\input{#2}}}{}%
}%
\providecommand \typeout [0]{\immediate \write \m@ne }%
\providecommand \selectlanguage [0]{\@gobble}%
\providecommand \bibinfo [0]{\@secondoftwo}%
\providecommand \bibfield [0]{\@secondoftwo}%
\providecommand \translation [1]{[#1]}%
\providecommand \BibitemOpen[0]{}%
\providecommand \bibitemStop [0]{}%
\providecommand \bibitemNoStop [0]{.\EOS\space}%
\providecommand \EOS [0]{\spacefactor3000\relax}%
\providecommand \BibitemShut [1]{\csname bibitem#1\endcsname}%
\bibitem{Huefner}%
  \BibitemOpen
  \bibfield{author}{%
  \bibinfo {author} {\bibfnamefont{S.}~\bibnamefont{H{\"u}fner}},\ }%
  \emph{\bibinfo {title} {Photoelectron Spectroscopy - Principles and
  Applications}}\ (\bibinfo {publisher} {Springer-Verlag},\ \bibinfo {year}
  {1996})%
  \bibAnnoteFile{NoStop}{Huefner}%
\bibitem{Stoehr}%
  \BibitemOpen
  \bibfield{author}{%
  \bibinfo {author} {\bibfnamefont{J.}~\bibnamefont{St{\"o}hr}},\ }%
  \emph{\bibinfo {title} {NEXAFS Spectroscopy}}\ (\bibinfo {publisher}
  {Springer-Verlag},\ \bibinfo {year} {1992})%
  \bibAnnoteFile{NoStop}{Stoehr}%
\bibitem{UenoProgSurfSci}%
  \BibitemOpen
  \bibfield{author}{%
  \bibinfo {author} {\bibfnamefont{N.}~\bibnamefont{Ueno}}\ and\ \bibinfo
  {author} {\bibfnamefont{S.}~\bibnamefont{Kera}},\ }%
  \bibfield{journal}{%
  \bibinfo {journal} {Prog. Surf. Sci.}\ }%
  \textbf{\bibinfo {volume} {83}},\ \bibinfo {pages} {490} (\bibinfo {year}
  {2008})%
  \bibAnnoteFile{NoStop}{UenoProgSurfSci}%
\bibitem{KeraProgSurfSci}%
  \BibitemOpen
  \bibfield{author}{%
  \bibinfo {author} {\bibfnamefont{S.}~\bibnamefont{Kera}}, \bibinfo {author}
  {\bibfnamefont{H.}~\bibnamefont{Yamane}},\ and\ \bibinfo {author}
  {\bibfnamefont{N.}~\bibnamefont{Ueno}},\ }%
  \bibfield{journal}{%
  \bibinfo {journal} {Prog. Surf. Sci.}\ }%
  \textbf{\bibinfo {volume} {84}},\ \bibinfo {pages} {135} (\bibinfo {year}
  {2009})%
  \bibAnnoteFile{NoStop}{KeraProgSurfSci}%
\bibitem{HwangMatSciEn}%
  \BibitemOpen
  \bibfield{author}{%
  \bibinfo {author} {\bibfnamefont{J.}~\bibnamefont{Hwang}}, \bibinfo {author}
  {\bibfnamefont{A.}~\bibnamefont{Wan}},\ and\ \bibinfo {author}
  {\bibfnamefont{A.}~\bibnamefont{Kahn}},\ }%
  \bibfield{journal}{%
  \bibinfo {journal} {Mater. Sci. Engin. R}\ }%
  \textbf{\bibinfo {volume} {63}},\ \bibinfo {pages} {1} (\bibinfo {year}
  {2009})%
  \bibAnnoteFile{NoStop}{HwangMatSciEn}%
\bibitem{SchoellPRL}%
  \BibitemOpen
  \bibfield{author}{%
  \bibinfo {author} {\bibfnamefont{A.}~\bibnamefont{Sch\"oll}}, \bibinfo
  {author} {\bibfnamefont{Y.}~\bibnamefont{Zou}}, \bibinfo {author}
  {\bibfnamefont{L.}~\bibnamefont{Kilian}}, \bibinfo {author}
  {\bibfnamefont{D.}~\bibnamefont{H\"ubner}}, \bibinfo {author}
  {\bibfnamefont{D.}~\bibnamefont{Gador}}, \bibinfo {author}
  {\bibfnamefont{C.}~\bibnamefont{Jung}}, \bibinfo {author}
  {\bibfnamefont{S.~G.}\ \bibnamefont{Urquhart}}, \bibinfo {author}
  {\bibfnamefont{T.}~\bibnamefont{Schmidt}}, \bibinfo {author}
  {\bibfnamefont{R.}~\bibnamefont{Fink}},\ and\ \bibinfo {author}
  {\bibfnamefont{E.}~\bibnamefont{Umbach}},\ }%
  \bibfield{journal}{%
  \bibinfo {journal} {Phys. Rev. Lett.}\ }%
  \textbf{\bibinfo {volume} {93}},\ \bibinfo {pages} {146406} (\bibinfo {year}
  {2004})%
  \bibAnnoteFile{NoStop}{SchoellPRL}%
\bibitem{ScholzPRL}%
  \BibitemOpen
  \bibfield{author}{%
  \bibinfo {author} {\bibfnamefont{M.}~\bibnamefont{Scholz}}, \bibinfo {author}
  {\bibfnamefont{F.}~\bibnamefont{Holch}}, \bibinfo {author}
  {\bibfnamefont{C.}~\bibnamefont{Sauer}}, \bibinfo {author}
  {\bibfnamefont{M.}~\bibnamefont{Wiessner}}, \bibinfo {author}
  {\bibfnamefont{A.}~\bibnamefont{Sch\"oll}},\ and\ \bibinfo {author}
  {\bibfnamefont{F.}~\bibnamefont{Reinert}},\ }%
  \bibfield{journal}{%
  \bibinfo {journal} {Phys. Rev. Lett.}\ }%
  \textbf{\bibinfo {volume} {111}},\ \bibinfo {pages} {048102} (\bibinfo {year}
  {2013})%
  \bibAnnoteFile{NoStop}{ScholzPRL}%
\bibitem{BobischJChemPhys}%
  \BibitemOpen
  \bibfield{author}{%
  \bibinfo {author} {\bibfnamefont{C.}~\bibnamefont{Bobisch}}, \bibinfo
  {author} {\bibfnamefont{T.}~\bibnamefont{Wagner}}, \bibinfo {author}
  {\bibfnamefont{A.}~\bibnamefont{Bannani}},\ and\ \bibinfo {author}
  {\bibfnamefont{R.}~\bibnamefont{M{\"o}ller}},\ }%
  \bibfield{journal}{%
  \bibinfo {journal} {J. Chem. Phys.}\ }%
  \textbf{\bibinfo {volume} {119}},\ \bibinfo {pages} {9804} (\bibinfo {year}
  {2003})%
  \bibAnnoteFile{NoStop}{BobischJChemPhys}%
\bibitem{ChenApplPhysLett}%
  \BibitemOpen
  \bibfield{author}{%
  \bibinfo {author} {\bibfnamefont{W.}~\bibnamefont{Chen}}, \bibinfo {author}
  {\bibfnamefont{H.}~\bibnamefont{Huang}}, \bibinfo {author}
  {\bibfnamefont{S.}~\bibnamefont{Chen}}, \bibinfo {author}
  {\bibfnamefont{L.}~\bibnamefont{Chen}}, \bibinfo {author}
  {\bibfnamefont{H.~L.}\ \bibnamefont{Zhang}}, \bibinfo {author}
  {\bibfnamefont{X.~Y.}\ \bibnamefont{Gao}},\ and\ \bibinfo {author}
  {\bibfnamefont{A.~T.~S.}\ \bibnamefont{Wee}},\ }%
  \bibfield{journal}{%
  \bibinfo {journal} {Appl. Phys. Lett.}\ }%
  \textbf{\bibinfo {volume} {91}},\ \bibinfo {pages} {114102} (\bibinfo {year}
  {2007})%
  \bibAnnoteFile{NoStop}{ChenApplPhysLett}%
\bibitem{KasemannLangmuir}%
  \BibitemOpen
  \bibfield{author}{%
  \bibinfo {author} {\bibfnamefont{D.}~\bibnamefont{Kasemann}}, \bibinfo
  {author} {\bibfnamefont{C.}~\bibnamefont{Wagner}}, \bibinfo {author}
  {\bibfnamefont{R.}~\bibnamefont{Forker}}, \bibinfo {author}
  {\bibfnamefont{T.}~\bibnamefont{Dienel}}, \bibinfo {author}
  {\bibfnamefont{K.}~\bibnamefont{M{\"u}llen}},\ and\ \bibinfo {author}
  {\bibfnamefont{T.}~\bibnamefont{Fritz}},\ }%
  \bibfield{journal}{%
  \bibinfo {journal} {Langmuir}\ }%
  \textbf{\bibinfo {volume} {25}},\ \bibinfo {pages} {12569} (\bibinfo {year}
  {2009})%
  \bibAnnoteFile{NoStop}{KasemannLangmuir}%
\bibitem{HaemingPRB}%
  \BibitemOpen
  \bibfield{author}{%
  \bibinfo {author} {\bibfnamefont{M.}~\bibnamefont{H{\"a}ming}}, \bibinfo
  {author} {\bibfnamefont{M.}~\bibnamefont{Greif}}, \bibinfo {author}
  {\bibfnamefont{C.}~\bibnamefont{Sauer}}, \bibinfo {author}
  {\bibfnamefont{A.}~\bibnamefont{Sch{\"o}ll}},\ and\ \bibinfo {author}
  {\bibfnamefont{F.}~\bibnamefont{Reinert}},\ }%
  \bibfield{journal}{%
  \bibinfo {journal} {Phys. Rev. B}\ }%
  \textbf{\bibinfo {volume} {82}},\ \bibinfo {pages} {1} (\bibinfo {year}
  {2010})%
  \bibAnnoteFile{NoStop}{HaemingPRB}%
\bibitem{HaemingSurfSci}%
  \BibitemOpen
  \bibfield{author}{%
  \bibinfo {author} {\bibfnamefont{M.}~\bibnamefont{H{\"a}ming}}, \bibinfo
  {author} {\bibfnamefont{M.}~\bibnamefont{Greif}}, \bibinfo {author}
  {\bibfnamefont{M.}~\bibnamefont{Wießner}}, \bibinfo {author}
  {\bibfnamefont{A.}~\bibnamefont{Sch{\"o}ll}},\ and\ \bibinfo {author}
  {\bibfnamefont{F.}~\bibnamefont{Reinert}},\ }%
  \bibfield{journal}{%
  \bibinfo {journal} {Surf. Sci.}\ }%
  \textbf{\bibinfo {volume} {604}},\ \bibinfo {pages} {1619} (\bibinfo {year}
  {2010})%
  \bibAnnoteFile{NoStop}{HaemingSurfSci}%
\bibitem{StadtmullerPRL}%
  \BibitemOpen
  \bibfield{author}{%
  \bibinfo {author} {\bibfnamefont{B.}~\bibnamefont{Stadtm\"uller}}, \bibinfo
  {author} {\bibfnamefont{T.}~\bibnamefont{Sueyoshi}}, \bibinfo {author}
  {\bibfnamefont{G.}~\bibnamefont{Kichin}}, \bibinfo {author}
  {\bibfnamefont{I.}~\bibnamefont{Kr\"oger}}, \bibinfo {author}
  {\bibfnamefont{S.}~\bibnamefont{Soubatch}}, \bibinfo {author}
  {\bibfnamefont{R.}~\bibnamefont{Temirov}}, \bibinfo {author}
  {\bibfnamefont{F.~S.}\ \bibnamefont{Tautz}},\ and\ \bibinfo {author}
  {\bibfnamefont{C.}~\bibnamefont{Kumpf}},\ }%
  \bibfield{journal}{%
  \bibinfo {journal} {Phys. Rev. Lett.}\ }%
  \textbf{\bibinfo {volume} {108}},\ \bibinfo {pages} {106103} (\bibinfo {year}
  {2012})%
  \bibAnnoteFile{NoStop}{StadtmullerPRL}%
\bibitem{NeebJourElSpec}%
  \BibitemOpen
  \bibfield{author}{%
  \bibinfo {author} {\bibfnamefont{M.}~\bibnamefont{Neeb}}, \bibinfo {author}
  {\bibfnamefont{J.-E.}\ \bibnamefont{Rubensson}}, \bibinfo {author}
  {\bibfnamefont{M.}~\bibnamefont{Biermann}},\ and\ \bibinfo {author}
  {\bibfnamefont{W.}~\bibnamefont{Eberhardt}},\ }%
  \bibfield{journal}{%
  \bibinfo {journal} {J. Electron Spectrosc. and Relat. Phenom.}\ }%
  \textbf{\bibinfo {volume} {67}},\ \bibinfo {pages} {261 } (\bibinfo {year}
  {1994})%
  \bibAnnoteFile{NoStop}{NeebJourElSpec}%
\bibitem{OsborneJChemPhys}%
  \BibitemOpen
  \bibfield{author}{%
  \bibinfo {author} {\bibfnamefont{S.}~\bibnamefont{Osborne}}, \bibinfo
  {author} {\bibfnamefont{A.}~\bibnamefont{Ausmees}}, \bibinfo {author}
  {\bibfnamefont{S.}~\bibnamefont{Svensson}}, \bibinfo {author}
  {\bibfnamefont{A.}~\bibnamefont{Kivim{\"a}ki}}, \bibinfo {author}
  {\bibfnamefont{O.}~\bibnamefont{Sairanen}}, \bibinfo {author}
  {\bibfnamefont{A.~N.}\ \bibnamefont{de~Brito}}, \bibinfo {author}
  {\bibfnamefont{H.}~\bibnamefont{Aksela}},\ and\ \bibinfo {author}
  {\bibfnamefont{S.}~\bibnamefont{Aksela}},\ }%
  \bibfield{journal}{%
  \bibinfo {journal} {J. Chem. Phys.}\ }%
  \textbf{\bibinfo {volume} {102}} (\bibinfo {year} {1995})%
  \bibAnnoteFile{NoStop}{OsborneJChemPhys}%
\bibitem{SundinPRL}%
  \BibitemOpen
  \bibfield{author}{%
  \bibinfo {author} {\bibfnamefont{S.}~\bibnamefont{Sundin}}, \bibinfo {author}
  {\bibfnamefont{F.}~\bibnamefont{Kh.~Gel'mukhanov}}, \bibinfo {author}
  {\bibfnamefont{H.}~\bibnamefont{\AA{}gren}}, \bibinfo {author}
  {\bibfnamefont{S.~J.}\ \bibnamefont{Osborne}}, \bibinfo {author}
  {\bibfnamefont{A.}~\bibnamefont{Kikas}}, \bibinfo {author}
  {\bibfnamefont{O.}~\bibnamefont{Bj\"orneholm}}, \bibinfo {author}
  {\bibfnamefont{A.}~\bibnamefont{Ausmees}},\ and\ \bibinfo {author}
  {\bibfnamefont{S.}~\bibnamefont{Svensson}},\ }%
  \bibfield{journal}{%
  \bibinfo {journal} {Phys. Rev. Lett.}\ }%
  \textbf{\bibinfo {volume} {79}},\ \bibinfo {pages} {1451} (\bibinfo {year}
  {1997})%
  \bibAnnoteFile{NoStop}{SundinPRL}%
\bibitem{KimbergPRX}%
  \BibitemOpen
  \bibfield{author}{%
  \bibinfo {author} {\bibfnamefont{V.}~\bibnamefont{Kimberg}}, \bibinfo
  {author} {\bibfnamefont{A.}~\bibnamefont{Lindblad}}, \bibinfo {author}
  {\bibfnamefont{J.}~\bibnamefont{S\"oderstr\"om}}, \bibinfo {author}
  {\bibfnamefont{O.}~\bibnamefont{Travnikova}}, \bibinfo {author}
  {\bibfnamefont{C.}~\bibnamefont{Nicolas}}, \bibinfo {author}
  {\bibfnamefont{Y.~P.}\ \bibnamefont{Sun}}, \bibinfo {author}
  {\bibfnamefont{F.}~\bibnamefont{Gel'mukhanov}}, \bibinfo {author}
  {\bibfnamefont{N.}~\bibnamefont{Kosugi}},\ and\ \bibinfo {author}
  {\bibfnamefont{C.}~\bibnamefont{Miron}},\ }%
  \bibfield{journal}{%
  \bibinfo {journal} {Phys. Rev. X}\ }%
  \textbf{\bibinfo {volume} {3}},\ \bibinfo {pages} {011017} (\bibinfo {year}
  {2013})%
  \bibAnnoteFile{NoStop}{KimbergPRX}%
\bibitem{KjeldgaardPRB}%
  \BibitemOpen
  \bibfield{author}{%
  \bibinfo {author} {\bibfnamefont{L.}~\bibnamefont{Kjeldgaard}}, \bibinfo
  {author} {\bibfnamefont{T.}~\bibnamefont{K\"a\"ambre}}, \bibinfo {author}
  {\bibfnamefont{J.}~\bibnamefont{Schiessling}}, \bibinfo {author}
  {\bibfnamefont{I.}~\bibnamefont{Marenne}}, \bibinfo {author}
  {\bibfnamefont{J.~N.}\ \bibnamefont{O'Shea}}, \bibinfo {author}
  {\bibfnamefont{J.}~\bibnamefont{Schnadt}}, \bibinfo {author}
  {\bibfnamefont{C.~J.}\ \bibnamefont{Glover}}, \bibinfo {author}
  {\bibfnamefont{M.}~\bibnamefont{Nagasono}}, \bibinfo {author}
  {\bibfnamefont{D.}~\bibnamefont{Nordlund}}, \bibinfo {author}
  {\bibfnamefont{M.~G.}\ \bibnamefont{Garnier}}, \bibinfo {author}
  {\bibfnamefont{L.}~\bibnamefont{Qian}}, \bibinfo {author}
  {\bibfnamefont{J.-E.}\ \bibnamefont{Rubensson}}, \bibinfo {author}
  {\bibfnamefont{P.}~\bibnamefont{Rudolf}}, \bibinfo {author}
  {\bibfnamefont{N.}~\bibnamefont{M\aa{}rtensson}}, \bibinfo {author}
  {\bibfnamefont{J.}~\bibnamefont{Nordgren}},\ and\ \bibinfo {author}
  {\bibfnamefont{P.~A.}\ \bibnamefont{Br\"uhwiler}},\ }%
  \bibfield{journal}{%
  \bibinfo {journal} {Phys. Rev. B}\ }%
  \textbf{\bibinfo {volume} {72}},\ \bibinfo {pages} {205414} (\bibinfo {year}
  {2005})%
  \bibAnnoteFile{NoStop}{KjeldgaardPRB}%
\bibitem{HaemingChemPhysLett}%
  \BibitemOpen
  \bibfield{author}{%
  \bibinfo {author} {\bibfnamefont{M.}~\bibnamefont{H{\"a}ming}}, \bibinfo
  {author} {\bibfnamefont{L.}~\bibnamefont{Weinhardt}}, \bibinfo {author}
  {\bibfnamefont{A.}~\bibnamefont{Sch{\"o}ll}},\ and\ \bibinfo {author}
  {\bibfnamefont{F.}~\bibnamefont{Reinert}},\ }%
  \bibfield{journal}{%
  \bibinfo {journal} {Chem. Phys. Lett.}\ }%
  \textbf{\bibinfo {volume} {510}},\ \bibinfo {pages} {82 } (\bibinfo {year}
  {2011})%
  \bibAnnoteFile{NoStop}{HaemingChemPhysLett}%
\bibitem{RoccoJChemPhys}%
  \BibitemOpen
  \bibfield{author}{%
  \bibinfo {author} {\bibfnamefont{M.~L.~M.}\ \bibnamefont{Rocco}}, \bibinfo
  {author} {\bibfnamefont{M.}~\bibnamefont{Haeming}}, \bibinfo {author}
  {\bibfnamefont{D.~R.}\ \bibnamefont{Batchelor}}, \bibinfo {author}
  {\bibfnamefont{R.}~\bibnamefont{Fink}}, \bibinfo {author}
  {\bibfnamefont{A.}~\bibnamefont{Sch{\"o}ll}},\ and\ \bibinfo {author}
  {\bibfnamefont{E.}~\bibnamefont{Umbach}},\ }%
  \bibfield{journal}{%
  \bibinfo {journal} {J. Chem. Phys.}\ }%
  \textbf{\bibinfo {volume} {129}},\ \bibinfo {pages} {074702} (\bibinfo {year}
  {2008})%
  \bibAnnoteFile{NoStop}{RoccoJChemPhys}%
\bibitem{GraberSurfSci}%
  \BibitemOpen
  \bibfield{author}{%
  \bibinfo {author} {\bibfnamefont{T.}~\bibnamefont{Graber}}, \bibinfo {author}
  {\bibfnamefont{F.}~\bibnamefont{Forster}}, \bibinfo {author}
  {\bibfnamefont{A.}~\bibnamefont{Sch{\"o}ll}},\ and\ \bibinfo {author}
  {\bibfnamefont{F.}~\bibnamefont{Reinert}},\ }%
  \bibfield{journal}{%
  \bibinfo {journal} {Surf. Sci.}\ }%
  \textbf{\bibinfo {volume} {605}},\ \bibinfo {pages} {878} (\bibinfo {year}
  {2011})%
  \bibAnnoteFile{NoStop}{GraberSurfSci}%
\bibitem{SchoelJourElSpec}%
  \BibitemOpen
  \bibfield{author}{%
  \bibinfo {author} {\bibfnamefont{A.}~\bibnamefont{Sch{\"o}ll}}, \bibinfo
  {author} {\bibfnamefont{Y.}~\bibnamefont{Zou}}, \bibinfo {author}
  {\bibfnamefont{T.}~\bibnamefont{Schmidt}}, \bibinfo {author}
  {\bibfnamefont{R.}~\bibnamefont{Fink}},\ and\ \bibinfo {author}
  {\bibfnamefont{E.}~\bibnamefont{Umbach}},\ }%
  \bibfield{journal}{%
  \bibinfo {journal} {J. Electron Spectrosc. and Relat. Phenom.}\ }%
  \textbf{\bibinfo {volume} {129}},\ \bibinfo {pages} {1 } (\bibinfo {year}
  {2003})%
  \bibAnnoteFile{NoStop}{SchoelJourElSpec}%
\bibitem{FanoPhysRev}%
  \BibitemOpen
  \bibfield{author}{%
  \bibinfo {author} {\bibfnamefont{U.}~\bibnamefont{Fano}},\ }%
  \bibfield{journal}{%
  \bibinfo {journal} {Phys. Rev.}\ }%
  \textbf{\bibinfo {volume} {124}},\ \bibinfo {pages} {1866} (\bibinfo {year}
  {1961})%
  \bibAnnoteFile{NoStop}{FanoPhysRev}%
\bibitem{DavisPRB}%
  \BibitemOpen
  \bibfield{author}{%
  \bibinfo {author} {\bibfnamefont{L.~C.}\ \bibnamefont{Davis}}\ and\ \bibinfo
  {author} {\bibfnamefont{L.~A.}\ \bibnamefont{Feldkamp}},\ }%
  \bibfield{journal}{%
  \bibinfo {journal} {Phys. Rev. B}\ }%
  \textbf{\bibinfo {volume} {23}},\ \bibinfo {pages} {6239} (\bibinfo {year}
  {1981})%
  \bibAnnoteFile{NoStop}{DavisPRB}%
\bibitem{EvangelistaJChemPhys}%
  \BibitemOpen
  \bibfield{author}{%
  \bibinfo {author} {\bibfnamefont{F.}~\bibnamefont{Evangelista}}, \bibinfo
  {author} {\bibfnamefont{V.}~\bibnamefont{Carravetta}}, \bibinfo {author}
  {\bibfnamefont{G.}~\bibnamefont{Stefani}}, \bibinfo {author}
  {\bibfnamefont{B.}~\bibnamefont{Jansik}}, \bibinfo {author}
  {\bibfnamefont{M.}~\bibnamefont{Alagia}}, \bibinfo {author}
  {\bibfnamefont{S.}~\bibnamefont{Stranges}},\ and\ \bibinfo {author}
  {\bibfnamefont{A.}~\bibnamefont{Ruocco}},\ }%
  \bibfield{journal}{%
  \bibinfo {journal} {J. Chem. Phys.}\ }%
  \textbf{\bibinfo {volume} {126}},\ \bibinfo {pages} {124709} (\bibinfo {year}
  {2007})%
  \bibAnnoteFile{NoStop}{EvangelistaJChemPhys}%
\bibitem{LinaresJourPhysChemB}%
  \BibitemOpen
  \bibfield{author}{%
  \bibinfo {author} {\bibfnamefont{M.}~\bibnamefont{Linares}}, \bibinfo
  {author} {\bibfnamefont{S.}~\bibnamefont{Stafstr{\"o}m}}, \bibinfo {author}
  {\bibfnamefont{Z.}~\bibnamefont{Rinkevicius}}, \bibinfo {author}
  {\bibfnamefont{H.}~\bibnamefont{Ågren}},\ and\ \bibinfo {author}
  {\bibfnamefont{P.}~\bibnamefont{Norman}},\ }%
  \bibfield{journal}{%
  \bibinfo {journal} {J. Phys. Chem. B}\ }%
  \textbf{\bibinfo {volume} {115}},\ \bibinfo {pages} {5096} (\bibinfo {year}
  {2011})%
  \bibAnnoteFile{NoStop}{LinaresJourPhysChemB}%
\bibitem{FrancescoJPhysChemA}%
  \BibitemOpen
  \bibfield{author}{%
  \bibinfo {author} {\bibfnamefont{R.}~\bibnamefont{De~Francesco}}, \bibinfo
  {author} {\bibfnamefont{M.}~\bibnamefont{Stener}},\ and\ \bibinfo {author}
  {\bibfnamefont{G.}~\bibnamefont{Fronzoni}},\ }%
  \bibfield{journal}{%
  \bibinfo {journal} {J. Phys. Chem. A}\ }%
  \textbf{\bibinfo {volume} {116}},\ \bibinfo {pages} {2885} (\bibinfo {year}
  {2012})%
  \bibAnnoteFile{NoStop}{FrancescoJPhysChemA}%
\bibitem{PiancastelliJPhysB}%
  \BibitemOpen
  \bibfield{author}{%
  \bibinfo {author} {\bibfnamefont{M.~N.}\ \bibnamefont{Piancastelli}},
  \bibinfo {author} {\bibfnamefont{M.}~\bibnamefont{Neeb}}, \bibinfo {author}
  {\bibfnamefont{A.}~\bibnamefont{Kivim{\"a}ki}}, \bibinfo {author}
  {\bibfnamefont{B.}~\bibnamefont{Kempgens}}, \bibinfo {author}
  {\bibfnamefont{H.~M.}\ \bibnamefont{K{\"o}ppe}}, \bibinfo {author}
  {\bibfnamefont{K.}~\bibnamefont{Maier}}, \bibinfo {author}
  {\bibfnamefont{A.~M.}\ \bibnamefont{Bradshaw}},\ and\ \bibinfo {author}
  {\bibfnamefont{R.~F.}\ \bibnamefont{Fink}},\ }%
  \bibfield{journal}{%
  \bibinfo {journal} {J. Phys. B: At. Mol. Opt. Phys.}\ }%
  \textbf{\bibinfo {volume} {30}},\ \bibinfo {pages} {5677 } (\bibinfo {year}
  {1997})%
  \bibAnnoteFile{NoStop}{PiancastelliJPhysB}%
\bibitem{TravnikovaJourElSpec}%
  \BibitemOpen
  \bibfield{author}{%
  \bibinfo {author} {\bibfnamefont{O.}~\bibnamefont{Travnikova}}, \bibinfo
  {author} {\bibfnamefont{C.}~\bibnamefont{Miron}}, \bibinfo {author}
  {\bibfnamefont{M.}~\bibnamefont{B{\"a}ssler}}, \bibinfo {author}
  {\bibfnamefont{R.}~\bibnamefont{Feifel}}, \bibinfo {author}
  {\bibfnamefont{M.}~\bibnamefont{Piancastelli}}, \bibinfo {author}
  {\bibfnamefont{S.}~\bibnamefont{Sorensen}},\ and\ \bibinfo {author}
  {\bibfnamefont{S.}~\bibnamefont{Svensson}},\ }%
  \bibfield{journal}{%
  \bibinfo {journal} {J. Electron Spectrosc. and Relat. Phenom.}\ }%
  \textbf{\bibinfo {volume} {174}},\ \bibinfo {pages} {100 } (\bibinfo {year}
  {2009})%
  \bibAnnoteFile{NoStop}{TravnikovaJourElSpec}%
\bibitem{BozekChemPhys}%
  \BibitemOpen
  \bibfield{author}{%
  \bibinfo {author} {\bibfnamefont{J.}~\bibnamefont{Bozek}}, \bibinfo {author}
  {\bibfnamefont{S.}~\bibnamefont{Canton}}, \bibinfo {author}
  {\bibfnamefont{E.}~\bibnamefont{Kukk}},\ and\ \bibinfo {author}
  {\bibfnamefont{N.}~\bibnamefont{Berrah}},\ }%
  \bibfield{journal}{%
  \bibinfo {journal} {Chem. Phys.}\ }%
  \textbf{\bibinfo {volume} {289}},\ \bibinfo {pages} {149 } (\bibinfo {year}
  {2003})%
  \bibAnnoteFile{NoStop}{BozekChemPhys}%
\bibitem{KukkPRA}%
  \BibitemOpen
  \bibfield{author}{%
  \bibinfo {author} {\bibfnamefont{E.}~\bibnamefont{Kukk}}, \bibinfo {author}
  {\bibfnamefont{G.}~\bibnamefont{Snell}}, \bibinfo {author}
  {\bibfnamefont{J.~D.}\ \bibnamefont{Bozek}}, \bibinfo {author}
  {\bibfnamefont{W.-T.}\ \bibnamefont{Cheng}},\ and\ \bibinfo {author}
  {\bibfnamefont{N.}~\bibnamefont{Berrah}},\ }%
  \bibfield{journal}{%
  \bibinfo {journal} {Phys. Rev. A}\ }%
  \textbf{\bibinfo {volume} {63}},\ \bibinfo {pages} {062702} (\bibinfo {year}
  {2001})%
  \bibAnnoteFile{NoStop}{KukkPRA}%
\bibitem{RemmersPRA}%
  \BibitemOpen
  \bibfield{author}{%
  \bibinfo {author} {\bibfnamefont{G.}~\bibnamefont{Remmers}}, \bibinfo
  {author} {\bibfnamefont{M.}~\bibnamefont{Domke}}, \bibinfo {author}
  {\bibfnamefont{A.}~\bibnamefont{Puschmann}}, \bibinfo {author}
  {\bibfnamefont{T.}~\bibnamefont{Mandel}}, \bibinfo {author}
  {\bibfnamefont{C.}~\bibnamefont{Xue}}, \bibinfo {author}
  {\bibfnamefont{G.}~\bibnamefont{Kaindl}}, \bibinfo {author}
  {\bibfnamefont{E.}~\bibnamefont{Hudson}},\ and\ \bibinfo {author}
  {\bibfnamefont{D.~A.}\ \bibnamefont{Shirley}},\ }%
  \bibfield{journal}{%
  \bibinfo {journal} {Phys. Rev. A}\ }%
  \textbf{\bibinfo {volume} {46}},\ \bibinfo {pages} {3935} (\bibinfo {year}
  {1992})%
  \bibAnnoteFile{NoStop}{RemmersPRA}%
\bibitem{OjiJChemPhys}%
  \BibitemOpen
  \bibfield{author}{%
  \bibinfo {author} {\bibfnamefont{H.}~\bibnamefont{Oji}}, \bibinfo {author}
  {\bibfnamefont{R.}~\bibnamefont{Mitsumoto}}, \bibinfo {author}
  {\bibfnamefont{E.}~\bibnamefont{Ito}}, \bibinfo {author}
  {\bibfnamefont{H.}~\bibnamefont{Ishii}}, \bibinfo {author}
  {\bibfnamefont{Y.}~\bibnamefont{Ouchi}}, \bibinfo {author}
  {\bibfnamefont{K.}~\bibnamefont{Seki}}, \bibinfo {author}
  {\bibfnamefont{T.}~\bibnamefont{Yokoyama}}, \bibinfo {author}
  {\bibfnamefont{T.}~\bibnamefont{Ohta}},\ and\ \bibinfo {author}
  {\bibfnamefont{N.}~\bibnamefont{Kosugi}},\ }%
  \bibfield{journal}{%
  \bibinfo {journal} {J. Chem. Phys.}\ }%
  \textbf{\bibinfo {volume} {109}},\ \bibinfo {pages} {10409} (\bibinfo {year}
  {1998})%
  \bibAnnoteFile{NoStop}{OjiJChemPhys}%
\bibitem{AristovJChemPhys}%
  \BibitemOpen
  \bibfield{author}{%
  \bibinfo {author} {\bibfnamefont{V.~Y.}\ \bibnamefont{Aristov}}, \bibinfo
  {author} {\bibfnamefont{O.~V.}\ \bibnamefont{Molodtsova}}, \bibinfo {author}
  {\bibfnamefont{V.~V.}\ \bibnamefont{Maslyuk}}, \bibinfo {author}
  {\bibfnamefont{D.~V.}\ \bibnamefont{Vyalikh}}, \bibinfo {author}
  {\bibfnamefont{V.~M.}\ \bibnamefont{Zhilin}}, \bibinfo {author}
  {\bibfnamefont{Y.~A.}\ \bibnamefont{Ossipyan}}, \bibinfo {author}
  {\bibfnamefont{T.}~\bibnamefont{Bredow}}, \bibinfo {author}
  {\bibfnamefont{I.}~\bibnamefont{Mertig}},\ and\ \bibinfo {author}
  {\bibfnamefont{M.}~\bibnamefont{Knupfer}},\ }%
  \bibfield{journal}{%
  \bibinfo {journal} {J. Chem. Phys.}\ }%
  \textbf{\bibinfo {volume} {128}},\ \bibinfo {pages} {034703} (\bibinfo {year}
  {2008})%
  \bibAnnoteFile{NoStop}{AristovJChemPhys}%
\bibitem{StadlerNaPhys}%
  \BibitemOpen
  \bibfield{author}{%
  \bibinfo {author} {\bibfnamefont{C.}~\bibnamefont{Stadler}}, \bibinfo
  {author} {\bibfnamefont{S.}~\bibnamefont{Hansen}}, \bibinfo {author}
  {\bibfnamefont{I.}~\bibnamefont{Kr{\"o}ger}}, \bibinfo {author}
  {\bibfnamefont{C.}~\bibnamefont{Kumpf}},\ and\ \bibinfo {author}
  {\bibfnamefont{E.}~\bibnamefont{Umbach}},\ }%
  \bibfield{journal}{%
  \bibinfo {journal} {Nat. Phys.}\ }%
  \textbf{\bibinfo {volume} {5}},\ \bibinfo {pages} {153} (\bibinfo {year}
  {2009})%
  \bibAnnoteFile{NoStop}{StadlerNaPhys}%
\bibitem{ZiroffUnpub}%
  \BibitemOpen
  \bibinfo {howpublished} {J. Ziroff \textit{et al.}, W{\"u}rzburg, unpublished
  (2011)}%
  \bibAnnoteFile{NoStop}{ZiroffUnpub}%
\end{thebibliography}
\end{document}